\documentclass[12pt]{article}

\usepackage{epsfig,latexsym,amsfonts,amsmath,amsthm,amssymb,amsbsy,multirow,slashed,wasysym,textcomp,subfigure}

\def\be{\begin{equation}}
\def\ee{\end{equation}}
\def\bea{\begin{eqnarray}}
\def\eea{\end{eqnarray}}

\newcommand{\eal}[1]{\be \begin{aligned} #1 \end{aligned}\ee} 
\newcommand{\eqn}[1]{\be #1 \ee} 
\newcommand{\eqa}[1]{\bea  #1\eea} 

\newcommand{\beq}{\begin{eqnarray}}
\newcommand{\eeq}{\end{eqnarray}}

\usepackage{color}

\long\def\new#1\endnew{{\bf #1}}		\long\def\del#1\enddel{}

\def\call{{\cal{L}}}
\def\calh{{\cal{H}}}
\def\cale{{\cal{E}}}
\def\calb{{\cal{B}}}
\def\calf{{\cal{F}}}
\def\calo{{\cal{O}}}

\begin{document}
 
\begin{titlepage}

\begin{flushright}
{IPhT-T07/165}\\
\end{flushright}
\bigskip
\bigskip
\bigskip
\centering{\Large \bf Metastable Supertubes \\ \vspace{0.3cm} and non-extremal Black Hole Microstates \vspace{0.9cm}}\\
\bigskip
\bigskip
\centerline{{\bf Iosif Bena, Andrea Puhm, Bert Vercnocke}}
\bigskip
\centering{Institut de Physique Th\'eorique, \\ CEA Saclay, 91191 Gif sur Yvette, France\vspace{0.4cm}}
\bigskip
\bigskip
\centerline{{ iosif.bena@cea.fr,~andrea.puhm@cea.fr,~bert.vercnocke@cea.fr, } }
\bigskip
\bigskip

\begin{abstract}
 
{We study the dynamics of supertubes in smooth bubbling geometries with three charges and three dipole charges that can describe black holes, black rings and their microstates. We find the supertube Hamiltonian in these backgrounds and show that there exist metastable supertube configurations, that can decay into supersymmetric and non-supersymmetric ones via brane-flux annihilation. We also find stable non-supersymmetric configurations. Both the metastable and the stable non-supersymmetric configuration are expected to describe microstate geometries for non-extremal black holes, and we discuss the implication of their existence for the fuzzball proposal.}
\end{abstract}
\end{titlepage}

\section{Introduction}

Metastable vacua appear to be a putative feature of supersymmetric gauge theories \cite{Intriligator:2006dd}, and the construction of such vacua in string theory has been the subject of much study. The most direct way to obtain such vacua in supergravity is to put antibranes in a background that has brane charge dissolved in flux. 

This has been first done by Kachru, Pearson and Verlinde \cite{Kachru:2002gs}, who argued that probe anti-D3 branes in a smooth Klebanov-Strassler solution \cite{Klebanov:2000hb} are metastable, and can decay by annihilating against the positive D3 brane charge dissolved in flux via a process termed ``brane-flux annihilation.'' Similarly, Klebanov and Pufu considered anti-M2 branes in a smooth solution with M2 brane charge dissolved in fluxes \cite{Cvetic:2000db}, and found that in the probe approximation these also give rise to metastable vacua \cite{Klebanov:2010qs}. While there is at this point some uncertainty about the fate of these antibrane constructions when full backreaction is taken into account \cite{Bena:2009xk, Bena:2010gs, Bena:2011hz, Giecold:2011gw, Blaback:2011nz}, it is clear that placing probe antibranes in flux backgrounds is the most obvious starting point in hunting for metastable vacua. 

Our purpose in this paper is to apply this procedure to supersymmetric black hole microstate geometries and to argue that in the probe approximation antibranes give rise to metastable solutions that should be interpreted as microstates of non-extremal black holes. 

\bigskip 

As is well-known (see \cite{Bena:2007kg} for a review) there exist huge families of supersymmetric  solutions that have the same charge, mass and angular-momentum as a black hole with a macroscopically-large horizon area, and have no horizon or singularity. The singularity of the extremal BPS black hole (that can be thought of as coming from the singular brane sources) is resolved by a geometric transition that yields a geometry with many ``bubbles`` threaded by fluxes. The charge of these ``black hole microstate'' solutions is entirely dissolved in these fluxes. 

The physics of these solutions strongly supports the fact that the time-like singularity of extremal black holes is resolved by low-mass modes that correct the geometry on scales comparable to the horizon scale, and completely excise the region between the singularity and the horizon. This certainly disagrees with the standard textbook picture of extremal Reissner-Nordstr\"om  black holes, but on the other hand it is exactly identical to the way string theory resolves other timelike singularities\footnote{Essentially all the timelike singularities that we know how to resolve in string theory: LMM, Klebanov-Strassler, the enhan\c con, the D1-D5 system, Polchinski-Strassler and its generalizations, are resolved by low-mass modes that modify the physics at a large distance away from the singularity \cite{Lin:2004nb, Klebanov:2000hb, Johnson:2001wm, Lunin:2001fv, Polchinski:2000uf, Bena:2000zb}.  }. Moreover, there is recent evidence from numerical relativity studies of extremal and near-extremal black holes that the region inside the horizon is unstable, and the instability yields a final configuration with essentially no spacetime inside the horizon (see \cite{Marolf:2010nd} for an overview, and \cite{Poisson:1990eh,Brady:1995ni,Dafermos:2003wr} for earlier work).

\begin{figure}[ht!]
\centering{
\subfigure[Extremal RN spacetime.]{
\hspace{.08\textwidth}
\includegraphics[width=.22\textwidth]{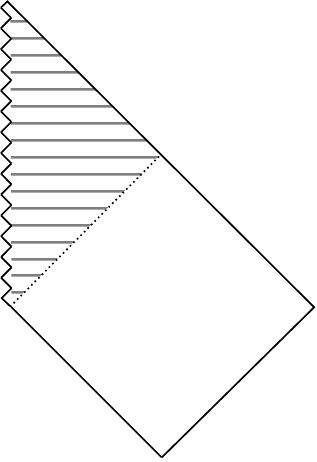}
\hspace{.08\textwidth}}       
\hspace{.04\textwidth}
\subfigure[Non-extremal RN spacetime.]{
\hspace{.08\textwidth}
\includegraphics[width=.32\textwidth]{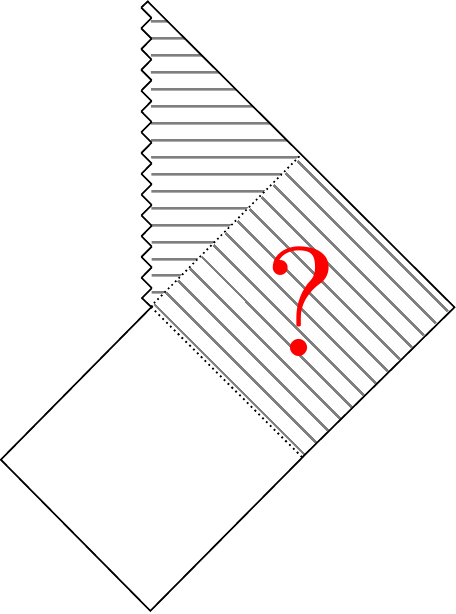}
\hspace{.08\textwidth}}
}
\caption{\small Penrose diagrams for the extremal (left) and non-extremal (right) Reissner-Nordstr\"om black hole. 
There is a growing body of work supporting the idea that the extremal black hole singularity is resolved at horizon scale.
For non-extremal black holes it is unclear whether the singularity resolution extends to the inner horizon, or all the way to the outer horizon, as in Mathur's proposal.
\label{fig:Penrose}}
\end{figure}

Hence, there appears to be a growing consensus that extremal black hole singularities are not resolved at Planck scale, but at horizon scale. While this fits very well with Mathur's conjecture\footnote{Also known as the fuzzball proposal, see \cite{Mathur:2005zp, Bena:2007kg, Mathur:2008nj, Balasubramanian:2008da, Skenderis:2008qn, Chowdhury:2010ct} for reviews.}, it leaves unanswered the question of what happens for non-extremal black holes. Indeed, given the resolution mechanism of the extremal black hole singularity, it is quite natural to expect that the timelike singularity of non-extremal black holes will be resolved on the scale of the inner horizon, and thus that there will be no more spacetime in the region from the singularity to the inner horizon; this also agrees with insights coming from numerical relativity \cite{Brady:1995ni}, and with ancient analytical work about the instability of inner horizons \cite{Penrose-Battelle, Poisson:1990eh,Dafermos:2003wr}.

Nevertheless, the fuzzball proposal, and in general the requirement that the information paradox be vindicated \cite{Mathur:2009hf} imply something much stronger: the singularity of non-extremal black holes should be resolved all the way to the outer horizon, in the past of the singularity. To understand whether this indeed happens, one must understand  what happens to the extremal black hole microstates when one tries to make them near-extremal. If they all collapse behind a horizon, then most likely the singularity resolution scale is that of the inner horizon, and these microstates do not help too much in solving the information paradox\footnote{Though there might exist microstate geometries that differ from the black hole solution all the way to the horizon, but still have a horizon \cite{Maloney}.}. On the other hand, if these microstates survive the near-extremal deformation and do not collapse behind a horizon, this indicates that the singularity is resolved all the way to the outer horizon, just like in the fuzzball proposal.

Note that settling the issue of the near-extremal fate of the microstate geometries is enough for arguing that the fuzzball proposal applies to all non-extremal black holes: the Penrose diagram of a large Schwarzschild black hole to which one adds one electron is the same as the one of the near-extremal Reissner-Nordstr\"om black hole, and if the singularity-resolution scale of the latter is the outer horizon, it is very implausible that it will be anything else for the former.

At this point only two classes of non-extremal microstate solutions are known: the JMaRT \cite{Jejjala:2005yu,Giusto:2007tt} and the running-Bolt \cite{Bena:2009qv, Bobev:2009kn} solutions, and all the arguments so far that the fuzzball proposal applies to non-extremal black holes have been based on the existence and physics of these solutions \cite{Cardoso:2005gj,Chowdhury:2007jx,Avery:2009tu}. Nevertheless, given that these solutions are very special, and given the technical difficulty involved in finding new ones, it is in rather difficult to take these arguments too far away.

Our purpose in this paper is to give a systematic method to construct and analyze near-extremal non-supersymmetric microstate solutions. We propose that starting from a generic extremal supersymmetric microstate solution one can build non-extremal microstates by adding certain branes with non-compatible supersymmetries. In particular we focus on two-charge supertubes (which correspond to fluxed D4 branes upon reduction to 4 dimensions), and find that a probe supertube can have both supersymmetric, as well as non-supersymmetric metastable minima. We explain the mechanism by which the supertube can annihilate against some of the charge dissolved in fluxes, and decay into a supersymmetric solution.

To prove the existence of metastable supertubes\footnote{We use the word supertube to refer to minima of the supertube Hamiltonian (or from a four-dimensional perspective to fluxed D4 branes), even when these configurations are not supersymmetric. This is because supertubes are locally-supersymmetric objects, and supersymmetry is broken because they are placed in a background of the wrong orientation.} and to understand their decay mechanism, it is enough to focus on supertubes in BPS microstate geometries built upon a Gibbons-Hawking base with two centers, and this is what we do in this paper. From the physics we find, it is quite obvious that metastable supertubes will exist in generic multi-center BPS three-charge geometries, and most likely also in non-BPS extremal multi-center solutions; we leave such generalizations for future work.

In Section 2 we compute the supertube Hamiltonian in an arbitrary three-charge solution with a Gibbons-Hawking base, and explore its minima analytically. In Section 3 we focus on a particular two-center background, and find that, depending on the supertube charges, it can have both supersymmetric and non-supersymmetric minima, as well as metastable minima. In Section 4 we describe how these metastable minima decay to the supersymmetric ones via brane-flux annihilation, and in Section 5 we discuss our results and future directions.
\medskip

{\bf Note:} When this article was in the final stages of preparation, the preprint \cite{Anninos:2011vn} appeared, which investigates the physics of probe metastable branes near single non-extremal black holes. The explicit system studied there is different from ours: we explore metastable configurations that probe {\it  multi-center} supersymmetric solutions and decay into supersymmetric or non-supersymmetric solutions via brane-flux annihilation. In \cite{Anninos:2011vn} the metastable branes probe {\it single-center} non-extremal black holes, and decay by falling behind the horizon. However, the physical conclusions of the two investigations point in the same direction: in the probe approximation there exist very large numbers of non-supersymmetric metastable configurations, that can be long-lived, and that play an important role in the physics of non-extremal black holes in string theory.

\section{The Supertube Hamiltonian}

The purpose of this section is to review the physics of supertubes in flat space, and to see how this extends when supertubes are placed in three-charge solutions constructed from a Gibbons-Hawking base, which descend in four dimensions to the multi-center solutions of \cite{Denef:2000nb,Bates:2003vx}.

\subsection{Supertubes in Flat Space}

A supertube is a brane configuration with two charges, a dipole charge, and angular momentum, that preserves 8 supersymmetries. In its original realization \cite{Mateos:2001qs,Mateos:2001pi, Emparan:2001ux}, the charges correspond to D0 branes and F1 strings dissolved into a tubular D2 brane. The tube has a non-trivial angular momentum in the space transverse to the strings which supports it from collapsing:
\be
|J|  = \Big|\frac{Q_{D0} Q_{F1}}{Q_{D2}}\Big| \,.\label{eq:RelationCharges}
\ee
To establish the existence of the supertube, one can examine the Born-Infeld action of the D2 brane with dissolved F1 and D0 charges, and find that the Hamiltonian of a tube of radius $R$  is 
\be
\calh = \frac{Q_{D2}}{R}\sqrt{Q_{D0}^2/Q_{D2}^2 +  R^2}\sqrt{Q_{F1}^2/Q_{D2}^2 + R^2}\label{eq:SupTube_FlatSpace_Ham}\ .
\ee
The Hamiltonian is minimized at
\be
R_{min} = \frac{\sqrt{|Q_{D0} Q_{F1}|}}{|Q_{D2}|}
\ee 
and is equal to the sum of the F1 and D0 charges. Thus, the minimum describes a supersymmetric tube.

\subsection{Tubes in three-charge backgrounds}

We now want to examine the dynamics of a supertube placed in a supersymmetric background geometry with three charges and three dipole charges, of the type that describe black holes, black rings and their microstate geometries. The metric in the M-theory duality frame in which the three charges correspond to M2 branes wrapping orthogonal $T^2$'s inside $T^6$ is \cite{Bena:2004de,Gutowski:2004yv}:
\be
ds_{11}^2 = (Z_1 Z_2 Z_3)^{-2/3}(dt + k)^2 + (Z_1 Z_2 Z_3)^{1/3} ds_4^2 
+(Z_1 Z_2 Z_3)^{1/3}\sum_I \frac{ds_I^2}{Z_I}
\label{11dgeometry}
\ee
where $ds_I^2$ are unit metrics on the three othogonal $T^2$'s and $ds_4^2 $ is the metric of the hyper-K\"{a}hler base space. When the latter metric is Gibbons-Hawking (GH) or Taub-NUT:
\bea
d s^2_4 &=& V^{-1} (d \psi + A)^2 + V ds_3^2\qquad \text{with} \qquad d A = \star_3 d V,
\eea
the solution is completely determined by specifying 8 harmonic functions $V,K^I,L_I,M$ in the GH base \cite{Gauntlett:2004qy,Elvang:2004ds}. 
In terms of these, the warp factors and rotation vector are given by
\bea
Z_I = L_I + \tfrac 12  C_{IJK}V^{-1} K^J K^K \qquad &\text{with}& \qquad C_{IJK}=|\epsilon_{IJK}|\,, \\
k = \mu (dt + \omega) + A \qquad  &\text{with}& \qquad \mu =  \tfrac 16 C_{IJK}V^{-2} K^I K^J K^K +\tfrac  12 L_I K^I + M\,,
\eea
where $ds_3^2$ is the flat metric on $\mathbb{R}^3$. The full details of the background geometry, including the background four-form flux, are given in appendix \ref{app:three_charge}. Note that the inverse of the warp factors $Z_I$ are also the electric potentials for the four-form and hence they determine the M2 charges at each background center. The two charges $Q_1$ and $Q_2$ of the supertube are parallel to those of the background and correspond to M2 branes along the first and second $T^2$. The dipole charge, $d_3$, corresponds to an M5 brane extended along those two tori wrapping the fiber of the Gibbons-Hawking space. If the tube is supersymmetric, then the fully-backreacted solution is again in the class  (\ref{11dgeometry}); the warp factors $Z_1$ and $Z_2$ will have a singularity at the location of the tube, and the Born-Infeld description of the supertube captures all the aspects of the backreacted solution \cite{Bena:2008dw}.

\begin{figure}[ht!]{ \centering{
\includegraphics[width=0.6\textwidth]{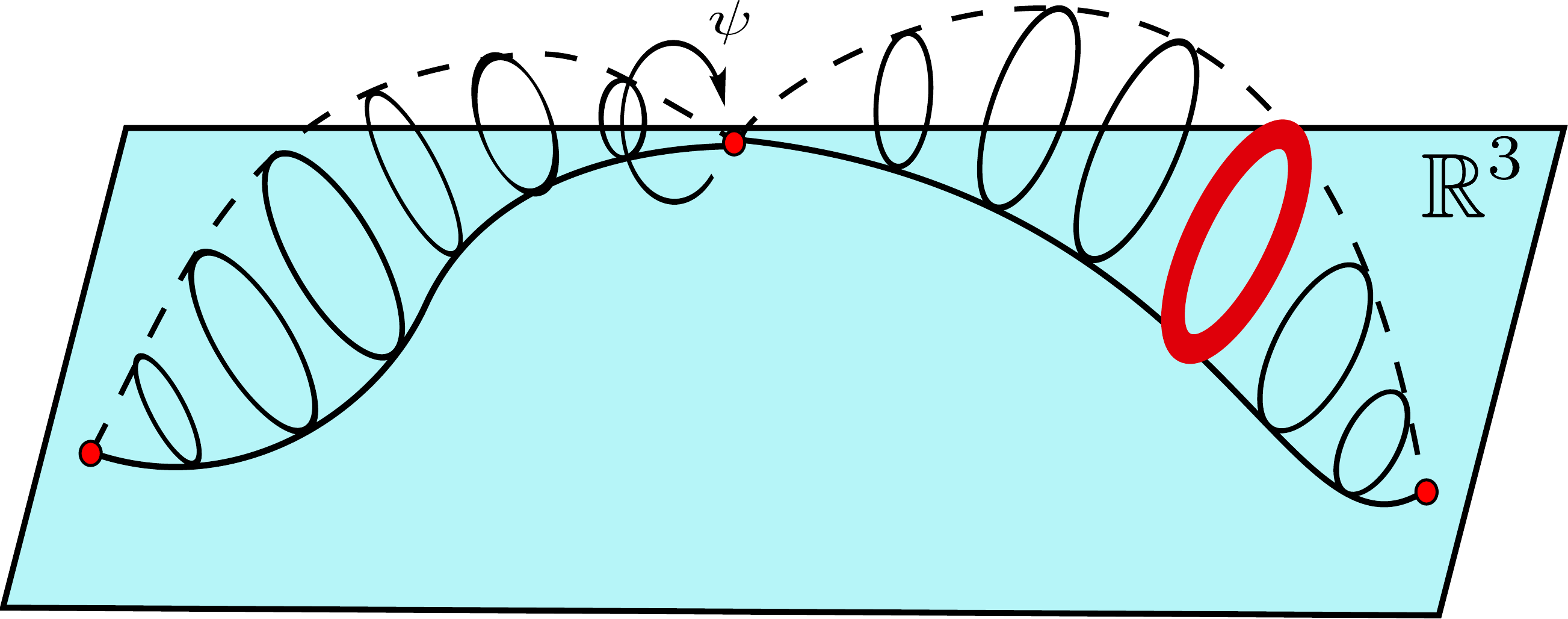}
} 
\caption{\small A smooth three-charge bubbling geometry with a supertube (red) placed on one of the cycles along $\psi$.}
\label{fig:bubblinggeometry}}
\end{figure}

However, since we are trying to study {\it non-supersymmetric} supertubes, for which no fully-backreacted description has been constructed so far, we will work in a probe approximation, ignoring the backreaction of the supertube. This is best done in a duality frame where the dynamics of the supertube can be described by a Born-Infeld action, and such a frame is obtained for example by reducing the 11-dimensional system along a torus direction. \medskip

We find that the Hamiltonian of a two-charge supertube in a multi-center three-charge background with Gibbons-Hawking base is:\footnote{See appendix \ref{app:Hamiltonian} for the derivation.}
\eal{
\calh &= \frac{\sqrt{Z_1Z_2Z_3 V^3}}{d_3(Z_1 Z_2Z_3 V - \mu^2 V^2)}\sqrt{\tilde Q_1^2 +d_3^2 \frac{Z_1Z_2Z_3 V - \mu^2 V^2}{Z_2^2 V^2}}\sqrt{\tilde Q_2^2 + d_3^2\frac{Z_1Z_2Z_3 V - \mu^2 V^2}{Z_1^2 V^2}}\\ &
+ \frac{\mu V^2}{d_3(Z_1Z_2Z_3V - \mu^2 V^2)}\tilde Q_1 \tilde Q_2 - \frac 1 {Z_1} \tilde Q_1 - \frac 1 {Z_2} \tilde  Q_2 -  \frac{ d_3 \mu}{Z_1Z_2} + Q_1 + Q_2 \,,\label{eq:Hamiltonian}
}
where we have introduced 
\be
\tilde Q_1 \equiv  Q_1 + d_3  (K^2/V - \mu/Z_2)~,~~~\tilde Q_2 \equiv  Q_2 + d_3  (K^1/V - \mu/Z_1)\,
\ee
and the harmonic functions $K^1$ and $K^2$ encode two of the three dipole moments of the background. Although the calculation yielding this result treats the two charges of the supertube differently, the final Hamiltonian is symmetric under interchange of indices ($1\leftrightarrow 2$), which is a non-trivial check. In flat space, the Hamiltonian \eqref{eq:Hamiltonian} reduces to \eqref{eq:SupTube_FlatSpace_Ham}: the first term becomes the  flat-space Hamiltonian and the terms in the second line vanish.

The combination $Z_1Z_2Z_3 V - \mu^2 V^2$ gives the square of the radius of the GH fiber $\psi$ and hence the square of the radius of the tube. In a regular background, free of closed timelike curves (CTC's), this combination is always positive, and hence the Hamiltonian is well-defined throughout the space. Furthermore, one can show that this Hamiltonian is always larger than or equal to  the sum of the supertube charges $Q_1$ and $Q_2$; for supersymmetric minima the equality comes about after a few non-trivial cancelations. 
The Hamiltonian \eqref{eq:Hamiltonian} describes tubes in the most general three-charge background. We expect this Hamiltonian to have both supersymmetric and non-supersymmetric minima.

An important difference between the flat space \eqref{eq:SupTube_FlatSpace_Ham} and the GH supertube Hamiltonian  \eqref{eq:Hamiltonian} is that the charges appearing in the latter are shifted from  the actual brane charges by a term proportional to the background magnetic fluxes\footnote{This shift is the crucial ingredient in the supertube entropy enhancement mechanism \cite{Bena:2008nh}.}.

\subsection{Minima of the Supertube Hamiltonian}

\paragraph{Supersymmetric minima.} The supersymmetric minima of the Hamiltonian have 
\be
V_{BPS}=Q_1 +Q_2 \,.
\ee
This value is obtained when the supertube radius is related to the supertube charges by
\be
d_3^2 \frac{Z_3} V =  (Q_1 + d_3\frac{K^2}{V}) (Q_2 + d_3\frac{K^1}V) \label{eq:Cond_SUSY}
\ee
and provided that 
\be
(Q_1 + d_3\frac{K^2}{V}) (Q_2 + d_3\frac{K^1}V) \geq0 \,.\label{eq:Bounds_SUSY1}
\ee
For this radius the Hamiltonian \eqref{eq:Hamiltonian} reduces to 
\be
\calh = Q_1 + Q_2 + \frac{Z_1 Z_2 Z_3 V}{Z_1 Z_2 Z_3 V - \mu^2 V^2} \Big[\ \Big|\frac{\tilde Q_1}{Z_1} + \frac{\tilde Q_2}{Z_2}\Big|- \Big(\frac{\tilde Q_1}{Z_1} + \frac{\tilde Q_2}{Z_2}\Big) \Big]
\ee
which saturates the BPS bound only if 
\be
\frac{\tilde Q_1}{Z_1} + \frac{\tilde Q_2}{Z_2}\geq0\,.\label{eq:Bounds_SUSY2}
\ee
When the conditions \eqref{eq:Bounds_SUSY1} and \eqref{eq:Bounds_SUSY2} are not met, the Hamiltonian has no supersymmetric minimum. 
The first condition \eqref{eq:Bounds_SUSY1} is needed for the absence of
 CTC's:  $Z_I V\geq 0$ (see appendix \ref{app:three_charge}, eq.\ \eqref{eq:CTC_cond}) implies that both sides of \eqref{eq:Cond_SUSY} have to be positive, which is equivalent to \eqref{eq:Bounds_SUSY1}. Similarly, if one is to construct the backreaction of the supertube, the warp factors near the supertube center (which we can take at $r=0$) would diverge \cite{Bena:2008dw} as $\left(Q_{1,2} + d_3 {K^{2,1}}/V\right)/r$. These divergences are controlled by the effective, or enhanced M2 charges
\be
Q_1^{\rm eff} \equiv Q_1 + d_3 \frac {K^2} V \qquad \text{and} \qquad  Q_2^{\rm eff} \equiv Q_2 + d_3 \frac {K^1} V\,.
\ee
If these charges do not have the same sign the solution has CTC's.

\paragraph{Non-supersymmetric minima.}

Given the complicated nature of the Hamiltonian (\ref{eq:Hamiltonian}), and given that we are trying to construct metastable black hole microstates, we focus from now on on supertubes in smooth bubbling multi-center solutions. The most generic such solution has a multi-center Taub-NUT or GH four-dimensional base space, with three non-trivial fluxes on the two-cycles  stretching between every pair of Taub-NUT or GH centers \cite{Bena:2005va,Berglund:2005vb,Saxena:2005uk}.

In order to study the existence of non-supersymmetric minima, we first expand the Hamiltonian near one of the smooth centers:
\eal{
\calh|_{r_i\to 0} 
 &= \Big|\Big(Q_1 + d_3\frac{K^2}{V}\Big)\Big(Q_2 +d_3\frac{K^1}{V}\Big)\Big| \sqrt{\frac{v_i}{Z_1 Z_2 Z_3}} \; \,(r_i)^{-1/2}  + \calo [(r_i)^0]\,,
}
where $r_i$ is the distance to the $i^{\rm th}$ center and $v_i$ is the coefficient of the $1/r_i$ pole in $V$. 

When both $(Q_1 +d_3K^2/V)|_{r_i=0}$ and $(Q_2 +d_3K^1/V)|_{r_i=0}$ are non-zero this Hamiltonian diverges near the centers\footnote{From a four-dimensional perspective, these supertubes are fluxed D4 branes with two non-zero fluxes, and hence a non-zero D0 charge; these are repelled by the fluxed D6 centers.} and hence there is at least one minimum between the centers. If  \eqref{eq:Bounds_SUSY1} is satisfied, the minimum is supersymmetric. To find a non-supersymmetric minimum one simply has to find a set of supertube charges such that at the minimum $(Q_1 +d_3K^2/V)(Q_2 +d_3K^1/V)<0$.

When one of the effective charges vanishes, say $(Q_1 +d_3 K^2/V)|_{r_i \rightarrow 0}=0$, then the supertube radius is zero. The divergent term of the Hamiltonian vanishes, and the leading term is:
\be
\calh|_{r_i\to 0} = Q_1 + Q_2 + \left|\frac{Q_2+d_3K^1/V}{Z_2}\right|- \frac{Q_2+d_3K^1/V}{Z_2} \,.
\label{eq:Degenerate_Hamiltonian}
\ee
One might naively think that a supertube of zero radius is nothing but a collection of branes, and may wonder why the dipole charge $d_3$ still appears in the Hamiltonian. The answer has to do with the existence of Dirac strings for the gauge field $A^{(1)}$ (given by eqs.\ \eqref{eq:11d_solution_app} and \eqref{eq:Magnetic_field} in appendix A). Since the solution has non-trivial fluxes, to completely describe the physics one must use multiple patches. The values of the gauge field $A^{(1)}$ differ from patch to patch, and in a generic patch there will be Dirac strings at generic centers. In particular, as we will discuss in detail in section 4, when $K^1/V$ is non-zero at the center, the zero-sized supertube described by \eqref{eq:Degenerate_Hamiltonian} wraps a Dirac string, and is not just a collection of parallel branes. When removing the Dirac string, the supertube brane charges shift, and become equal to the effective charges. 

It is then the orientation of these effective charges with respect to the background that determines whether supersymmetry is broken or not. In our example, when the effective charge of the supertube, $Q_2+d_3K^1/V$, has the same orientation as the charge of the background (proportional to $Z_2$), the Hamiltonian is equal to the sum of the charges, and the configuration is supersymmetric.
When this effective charge has the opposite orientation, the Hamiltonian is strictly larger than $V_{BPS}$, and the configuration is a non-supersymmetric minimum or maximum (depending on the sign of the next-to-leading order term).

\medskip 

Given that supersymmetric tubes have a critical worldvolume electric field, one may attempt to analytically obtain  a non-supersymmetric minimum using a supertube with a critical electric field oriented opposite to the background ($(Q_2+d_3\frac{K^1}{V})/Z_2 r_i<0$). When one of the effective charges is zero, this naive guess yields a non-supersymmetric minimum with energy 
\be
 V_{non-BPS} = Q_1 + Q_2 -2\frac{Q_2+d_3 {K^1}/{V}}{Z_1}\,. 
\ee
which agrees with \eqref{eq:Degenerate_Hamiltonian}! Thus the naive guess gives the correct energy of the zero-radius non-supersymmetric configuration. 
The naive guess also gives a radius relation: 
\be
-\frac{Z_3}{V}=(Q_1+d_3\frac{K^2}{V}-2d_3\frac{\mu}{Z_2})(Q_2+d_3\frac{K^1}{V})\,,
\label{naive-radius}
\ee
which is satisfied trivially for degenerate supertubes.
Unfortunately, non-degenerate supertubes do not have a critical electric field, and the naive minima obtained from \eqref{naive-radius} do not agree with those of the Hamiltonian. It would be interesting to get a deeper understanding of why this naive guess describes correctly the metastable vacua with degenerate supertubes but not the other ones. 

To summarize this section, one can use the Hamiltonian (\ref{eq:Hamiltonian}) to infer analytically the existence and properties of degenerate vacua, as well as the existence of non-degenerate vacua. However, we could not find any easy analytic way to describe non-degenerate non-supersymmetric vacua. Thus, we will now focus on a simple two-center bubbling solution, and analyze the possible minima numerically.

\section{Metastable supertubes in a two-center solution}

In this section, we evaluate the Hamiltonian in a specific two-center solution. Depending on the values of their charges $Q_1,Q_2$, the supertubes can have a rich structure of minima and interesting decay patterns. We scan for the candidate charges yielding metastable vacua by first plotting the supertube Hamiltonian on the axis between the centers.
We find examples with either one supersymmetric, or one non-supersymmetric, stable minimum. There can also be two minima for a given set of supertube charges $Q_1,Q_2$.
Either they are both supersymmetric or one is stable (supersymmetric or non-supersymmetric) and the other is metastable. We expect this rich minima structure to carry over to other smooth background geometries with {\it multi}-center Taub-NUT base spaces. When the candidate configuration is  metastable, we also plot the Hamiltonian away from the axes to insure that the supertube is not unstable to rolling away from the symmetry axis. 

\subsection{Details of the background}

The Hamiltonian depends on the spacetime coordinates through the harmonic functions
\be
\begin{aligned}
V &= v_0 + \frac{v_1}{|\vec r - \vec r_1|}+ \frac{v_2}{|\vec r - \vec r_2|}\,, \qquad &M &= m_0 + \frac{m_1}{|\vec r - \vec r_1|}+ \frac{m_2}{|\vec r - \vec r_2|}\,,\\ 
K^I &= k^I_0 + \frac{k^I_1}{|\vec r - \vec r_1|}+ \frac{k^I_2}{|\vec r - \vec r_2|}\,,  &L_I &= \ell_{I,0} + \frac{\ell_{I,1}}{|\vec r - \vec r_1|}+ \frac{\ell_{I,2}}{|\vec r - \vec r_2|}\,.
\end{aligned}
\ee
For an interpretation of the charges see table \ref{tab:IIA_Charges} in appendix \ref{app:three_charge}.
We choose a cylindrical coordinate system $(\rho,z,\theta)$ in three dimensions, where $z$ runs along the axis through the center and $\rho,\theta$ are polar coordinates in the orthogonal plane.  Since we have cylindrical symmetry, the solution only depends on the coordinates $z$ and $\rho$. 

We consider a two-center background where the inter-center distance is $r_{12} =|\vec r_1 - \vec r_2|= 120$ and we fix the two centers on the symmetry axis as $z_1 = -60, z_2 = 60$. We choose $K^I = K$ all equal and the following charges and asymptotics in the harmonic functions $V,K$: 
\be
\begin{aligned}
 &v_0=2\,,&\quad &v_1= 10\,,\quad &v_2 = -2\,, \\
 &k_0=-3\,,&\quad &k_1=0\,,\quad&k_2=14\,.
\end{aligned}
\ee
Because at spatial infinity $V \rightarrow v_0 \neq 0$, the GH space is in fact a two-center Taub-NUT space asymptotic to $\mathbb{R}^3 \times S^1$. Furthermore, $K$ has a non-vanishing constant asymptotic value, and hence the solution has non-trivial Wilson lines along the Gibbons-Hawking fiber (which descend to axions upon reduction to four dimensions). This choice of charges and asymptotic values is a representative choice that has the generic features of a smooth geometry with a two-center Taub-NUT base space. Furthermore, with the choice $k_1 = 0$, there are no Dirac strings at center ``1''.

Note that the harmonic functions $M$ and $L_I$ are completely determined through the regularity conditions on $Z_I$ and $\mu$ and the bubble equations (eqs.\ \eqref{eq:regularityharm} and \eqref{eq:smoothbubble} in appendix \ref{app:three_charge}). Finally, without loss of generality we consider a tube with dipole charge $d_3=1$.

\subsection{Plots of the potential}

We plot the Hamiltonian, shifted to $\calh \to \calh - (Q_1 +Q_2)$ (such that $\calh =0$ for supersymmetric tubes), in terms of the coordinate $z$ at $\rho=0$ (i.e.\ on the symmetry axis). Varying the supertube charges, a rich structure of different minima arises:

\subsubsection{One minimum}
\begin{figure}[ht!]{
\centering
\subfigure[One supersymmetric minimum.  ]{
 \includegraphics[width=0.45\textwidth]{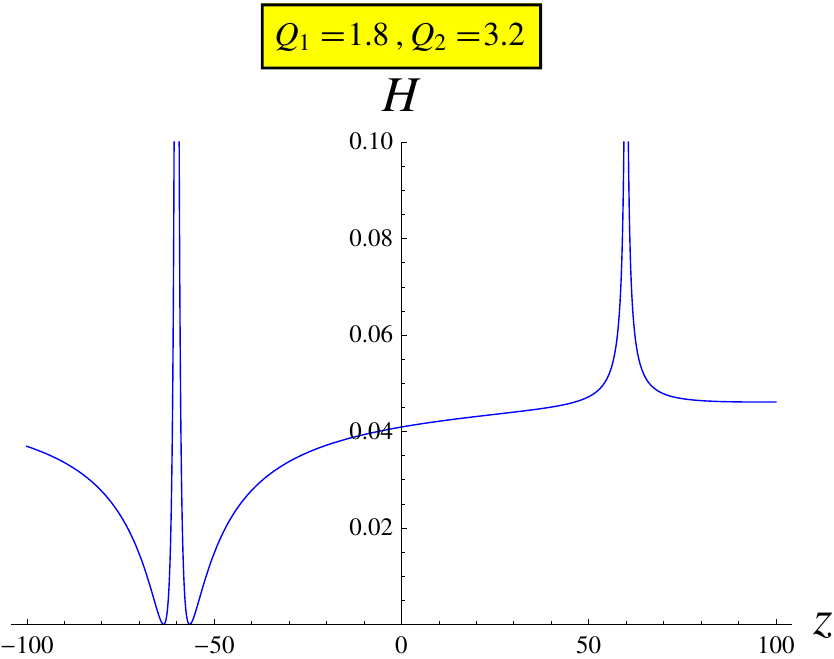}
\label{fig:1min_S}
}
\hspace{.05\textwidth}
\subfigure[One non-supersymmetric minimum.]{
 \includegraphics[width=.45\textwidth]{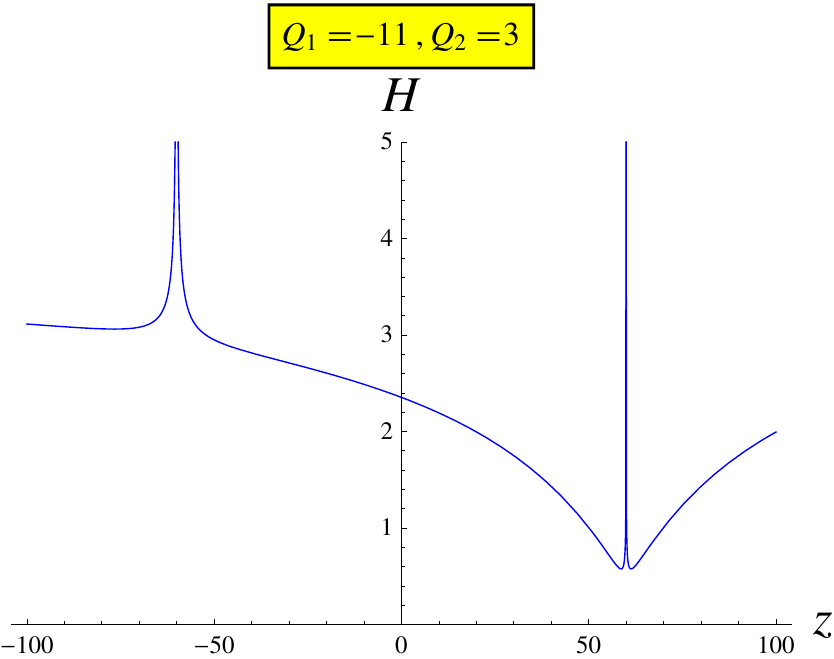}
\label{fig:1min_NS}}
\label{fig:singlemin}
}
\caption{\small{A single stable minimum between the two centers. When both charges are positive and have the same orientation as the background (which has electric potential $Z >0$ at the left center), the minimum is supersymmetric. When one of the charges has the wrong orientation, the minimum is non-supersymmetric. Note that the apparent second minimum outside the 2-center range is connected to the one inside by a Mexican-hat-type potential around the center in the $z-\rho$ plane. As we will show below in fig. \ref{fig:Mexican_Hat}, when supersymmetry is broken this Mexican-hat potential is slightly tilted.}}
\end{figure}

We first choose charges $Q_1,Q_2$ for which there is one minimum. When $\calh = 0$ the minimum is supersymmetric, (see figure \ref{fig:1min_S}), and the supertube generically sits close to one of the centers. The supersymmetric minimum describes a supertube whose radius can be found from eq.\ \eqref{eq:Cond_SUSY}, and the backreaction of this configuration is a supersymmetric solution with three centers \cite{Bena:2008dw}. 

 A minimum with $\calh \neq 0$ describes a  non-supersymmetric supertube (see figure \ref{fig:1min_NS}). Given that this tube is locally-BPS, we expect its backreaction to yield a solution that in the D1-D5-P duality frame is smooth at the supertube location. This smooth solution should represent a microstate of a non-supersymmetric black hole. It is interesting to ask whether the absolutely-stable non-supersymmetric minimum, which has no obvious decay channel, may correspond to an extremal non-BPS black hole, and hence fit in the recent classification of \cite{Bossard:2011kz}, or whether it will represent a very long-lived microstate of a non-extremal black hole.

\begin{figure}[htb!]
\centering{
\subfigure[Degenerate supersymmetric minimum.]{
\includegraphics[width=.45\textwidth]{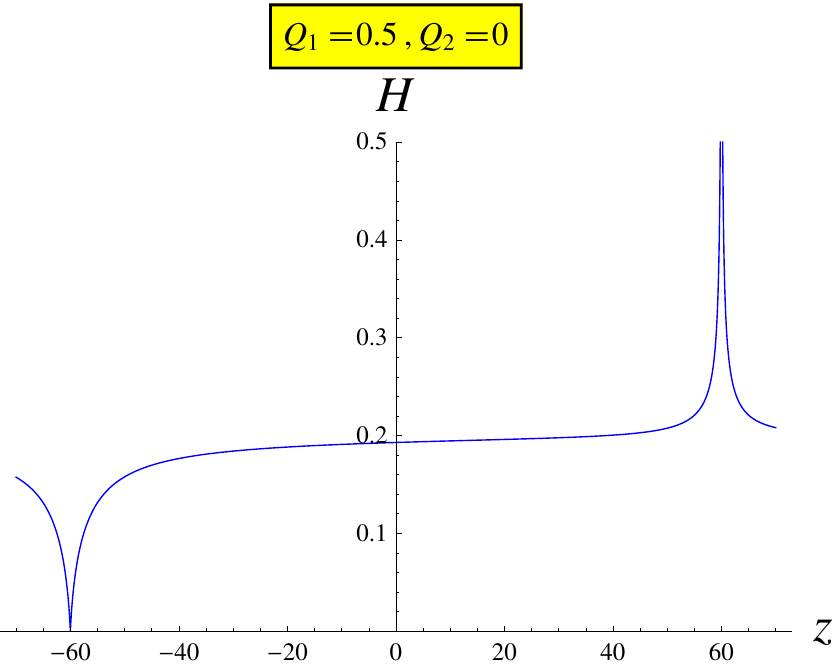}
\label{fig:1min_Sp}}
\hspace{.05\textwidth}
\subfigure[Degenerate non-supersymmetric minimum.]{
\includegraphics[width=.45\textwidth]{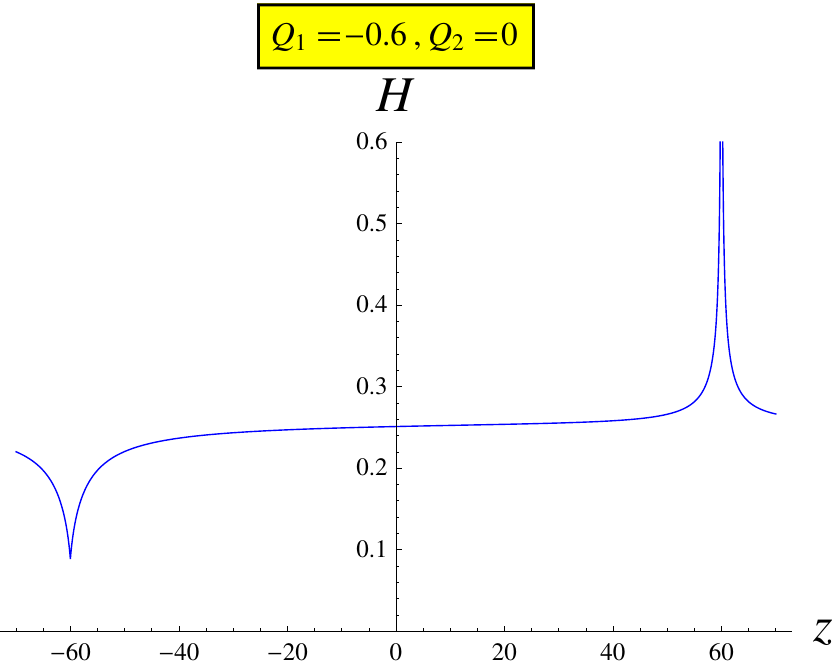}
\label{fig:1min_NSp}}
}
\caption{\small When one of the effective charges is zero, the supertube degenerates. In this example, we choose $Q_2 + \frac{K}{V}\big|_{r_1} = 0$. (Remember we work in patch ``1'' where $\frac{K}{V}\big|_{r_1}=0$.) When the other charge has the same orientation as the background, the minimum is supersymmetric (a), when the orientations are opposite, the minimum is non-supersymmetric (b).}
\end{figure}

As explained analytically in the previous section, for some choices of charges there also
exist supersymmetric and non-supersymmetric stable minima where the supertube has zero size. As expected, this happens when one of the \emph{effective} M2 charges $Q_I^{\rm eff} = Q_I + K/V$ is zero (see figures \ref{fig:1min_Sp} and \ref{fig:1min_NSp}).

\subsubsection{Two minima.}

\begin{figure}[ht!]{
\centering
\subfigure[Smooth supersymmetric minima.]{
\includegraphics[width=.45\textwidth]{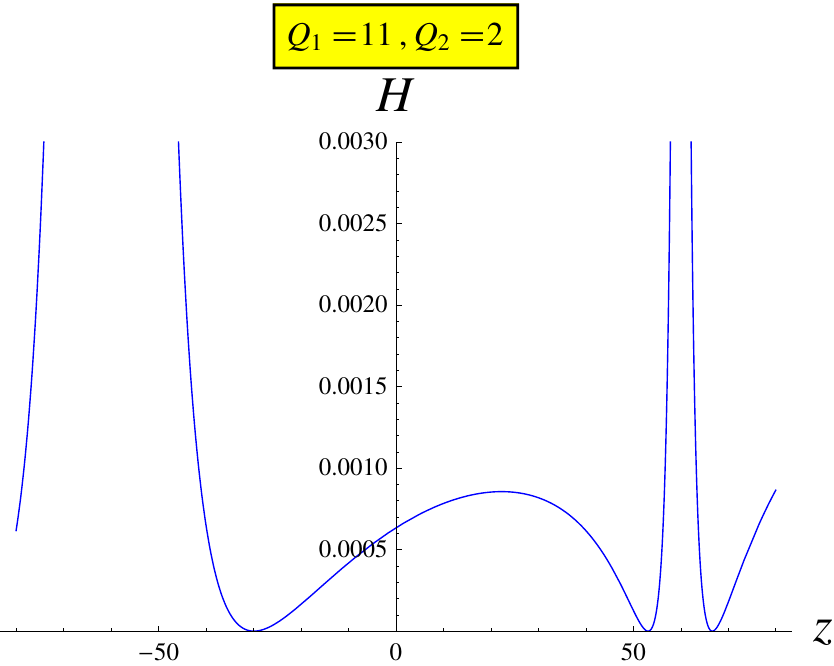} 
\label{fig:2min_S-Ssmooth}
} 
\hspace{.05\textwidth}
\subfigure[Smooth and degenerate supersymmetric minima.]{
 \includegraphics[width=.45\textwidth]{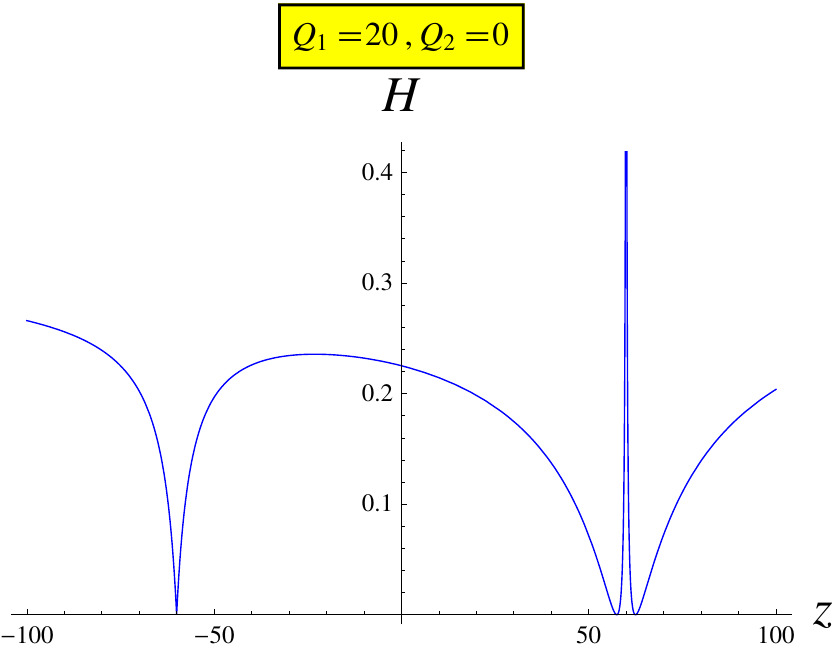}
\label{fig:2min_S-Sdegen}
}
\label{fig:2min_S-S}
\caption{\small{Two supersymmetric minima between the centers. When $Q_2 = 0$ in patch ``1'', the minimum close to the left center degenerates, and the supertube becomes a collection of parallel branes. Again, the additional minima outside the 2-center range are connected to the ones inside by a Mexican-hat-type potential around the respective center in the $z-\rho$ plane.}}}
\end{figure}

There is also a wide range of charges for which two minima appear. The first possibility is to have two supersymmetric minima, which is depicted in figure \ref{fig:2min_S-S}.

Figure \ref{fig:2min_S-Ssmooth} illustrates two different positions (each one close to one of the background centers) at which a supertube with fixed charges $Q_1,Q_2$ can be located. Again, one of these minima can degenerate when one of the effective charges $Q_I + K/V$ goes to zero (figure \ref{fig:2min_S-Sdegen}).  

\begin{figure}[ht!]{
\centering
\subfigure[Metastable and supersymmetric stable minima.]{
\includegraphics[width=.45\textwidth]{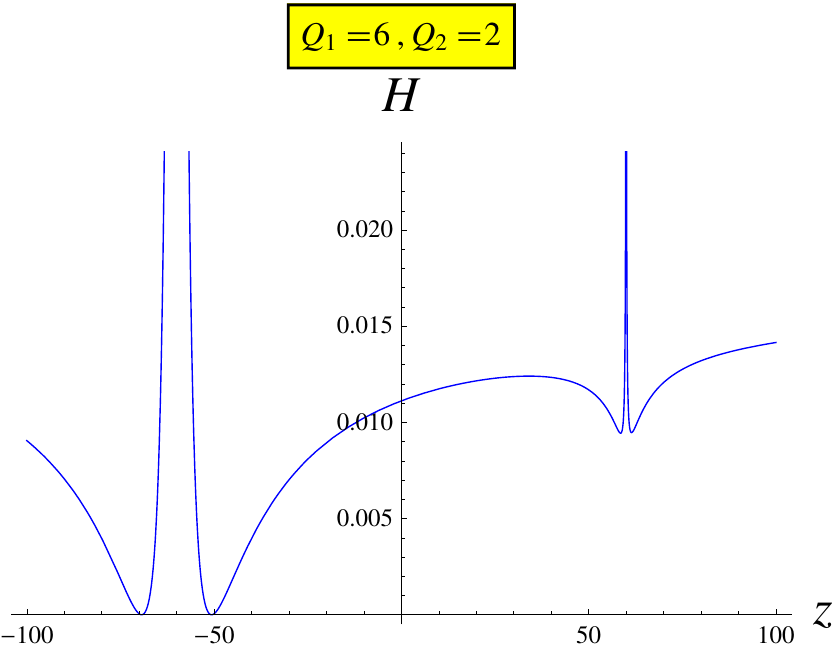}\label{fig:metastable_susy}}
\hspace{.05\textwidth}
\subfigure[Metastable and non-supersymmetric stable minima.]{
\includegraphics[width=.45\textwidth]{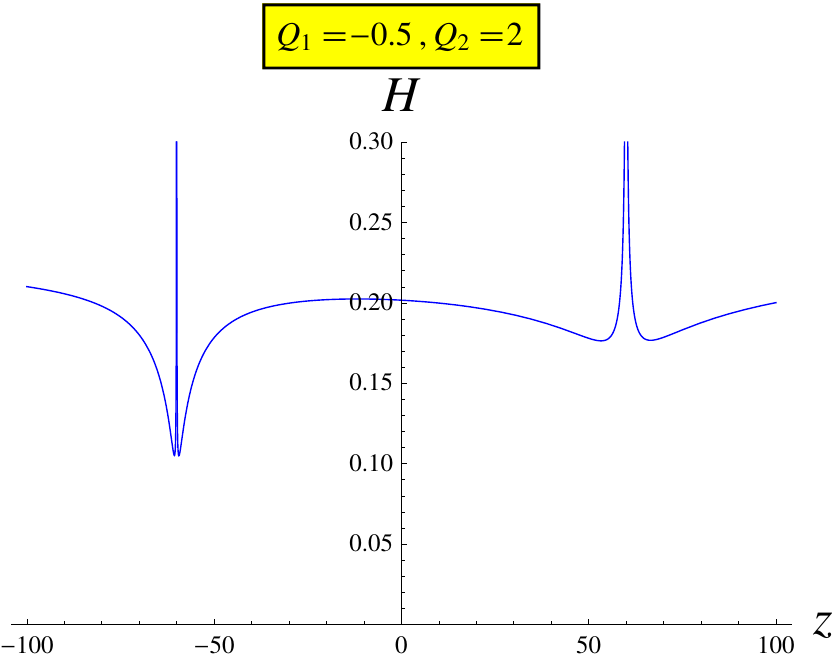}\label{fig:metastable_nonsusy}}\\
\subfigure[Contour Plot of \ref{fig:metastable_susy} in a plane of fixed polar angle around the symmetry axis (the $(z,\rho)$-plane): darker colours mean lower energy, the color scales in the main figure and the insets are not  the same.]{
\includegraphics[width=.4\textwidth]{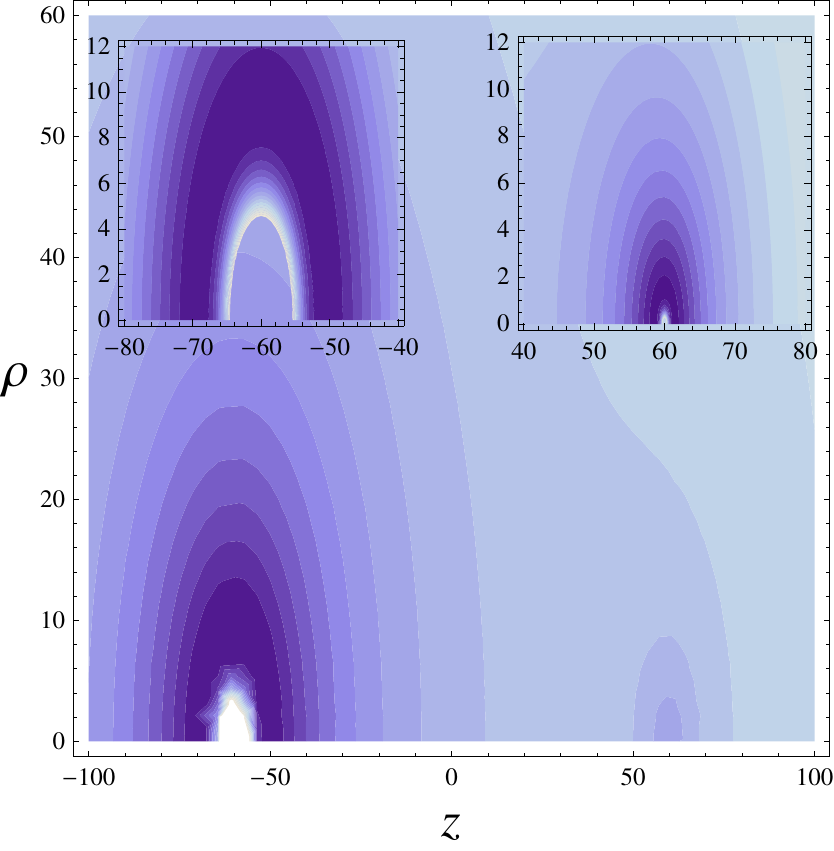}\label{fig:metastable_susy_contour}}
\hspace{.08\textwidth}
\subfigure[Contour Plot of \ref{fig:metastable_nonsusy} in the $(z,\rho)$-plane: darker colours mean lower energy, the color scales in the main figure and the insets are not  the same.]{
\includegraphics[width=.4\textwidth]{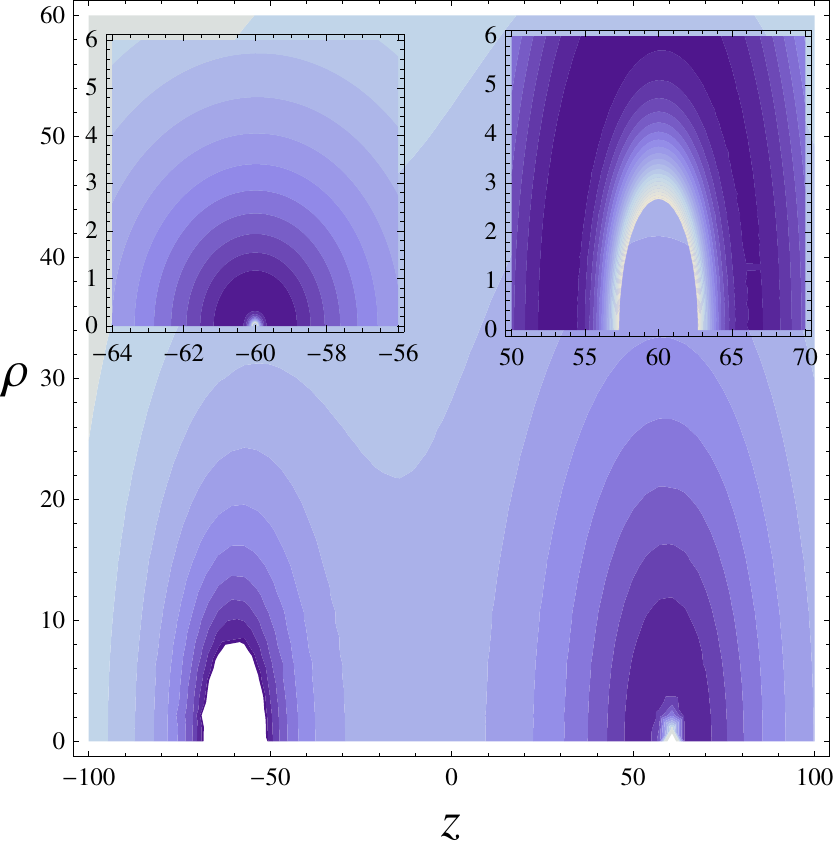}}
\caption{\small Metastable configurations.  The supertube charges are given in the patch where there are no Dirac strings at the left center (point ``1''). At this center, the M2 charge of the background is positive $( Z >0)$. When both charges are aligned with the background (left), the stable minimum is supersymmetric. When at least one of the supertube charges has the wrong orientation, the lowest minimum is non-supersymmetric.}
\label{fig:metastable}}
\end{figure}

The most interesting potentials arise when at least one of the two minima is non-supersymmetric, as depicted in figure \ref{fig:metastable}. These describe a metastable tube close to one center that can decay to a stable tube close to the other center. The plots in figure \ref{fig:metastable} show that the stable tube can be either supersymmetric or non-supersymmetric. 

\begin{figure}[ht!]
\centering
\subfigure{
\includegraphics[width=.3\textwidth]{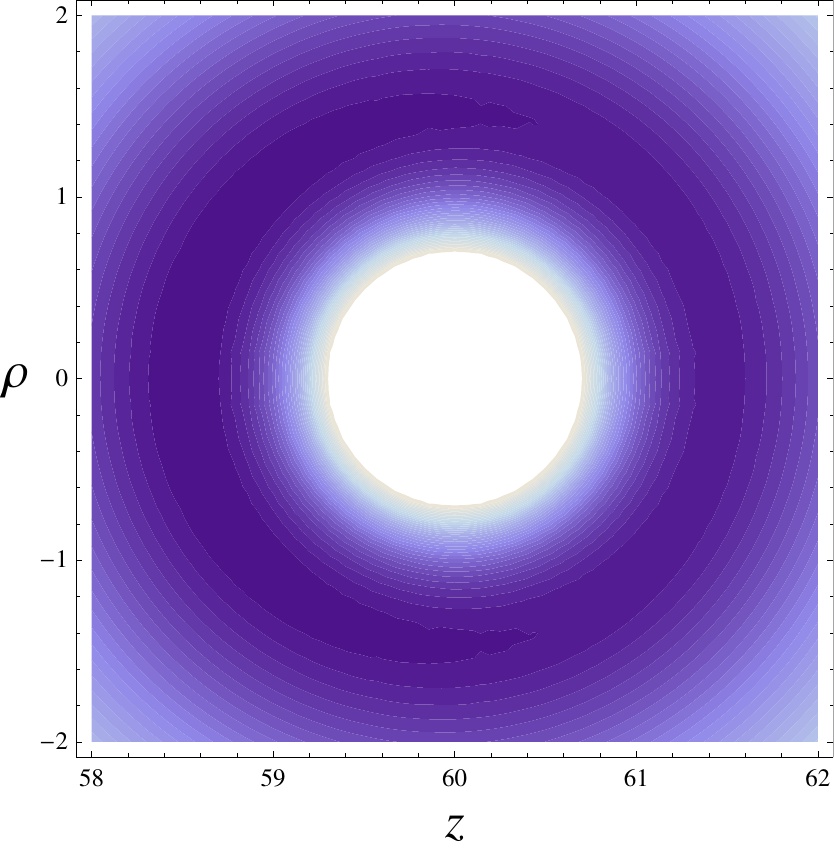}}
\hspace{.05\textwidth}
\subfigure{
\includegraphics[width=.3\textwidth]{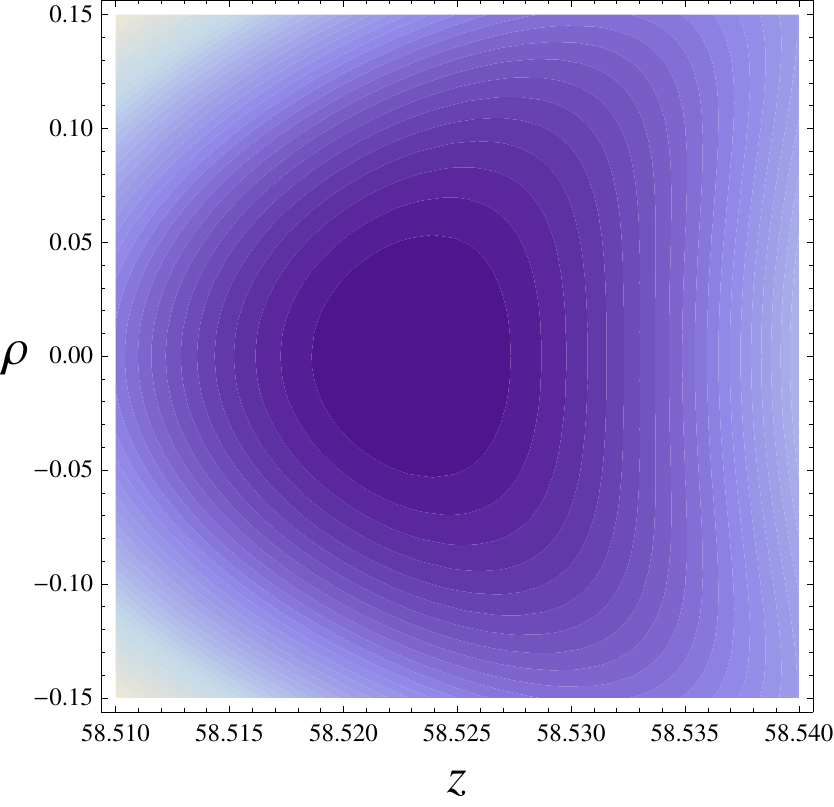}}
\caption{Two consecutives zooms on the minimum near $z = 60$ with $Q_1 = 6, Q_2 = 2$ of fig.\ \ref{fig:metastable_susy_contour}, shows the ``Mexican-hat'' type potential, which is slightly tilted to the left. (Darker colours mean lower energy.) The actual metastable minimum lies between the two centers on the symmetry axis. \label{fig:Mexican_Hat}}
\end{figure}

Note that near the metastable minimum the potential looks like a Mexican hat brim that is very slightly tilted. This is hard to see from figure \ref{fig:metastable}, and is shown in figure \ref{fig:Mexican_Hat}.

\section{The decay of metastable supertubes}

When the metastable supertube tunnels to the stable minimum, its quantized M2 charges stay the same, but its effective M2 charges
\be
Q_1^{\rm eff} =  Q_1 + d_3 \frac{K^2} V\,,\ \qquad Q_2^{\rm eff} =  Q_2 + d_3 \frac{K^1} V\,.
\ee 
change. Much like the decay of antibranes in backgrounds with charge dissolved in flux \cite{Kachru:2002gs,Klebanov:2010qs}, the decay of the metastable supertube can be understood as brane-flux annihilation. However, since the supertube has multiple charges, and generically has a non-zero radius both before and after the decay, the details are a bit more involved. 

\medskip 

Recall that the solution has a non-trivial magnetic flux through the two-cycle between the two GH points, given by 
\be
\Pi_{12}^{(I)}=\frac{1}{4\pi} \int_{r_1}^{r_2} dB^{(I)} = \frac{K^I}{V}\Big|_{r_2}-\frac{K^I}{V}\Big|_{r_1}\,.
\ee
Since the supertube has a non-trivial $d_3$ dipole charge which couples magnetically to $B^{(3)}$, when the supertube sweeps out the two-cycle from the North Pole to the South Pole, the amount of $\Pi_{12}^{(3)}$ flux on this two-cycle decreases by $d_3$ units. 

Imagine now lowering from infinity a supertube to the metastable minimum. If this minimum is, say, near the North Pole, in order to bring the supertube ``into position'' we need to work in a patch where there is no Dirac string at this pole. The change of patch in supergravity is realized by a gauge transformation \cite{Bena:2005ni,Bena:2007kg}, which transforms the eight harmonic functions via
\be
\begin{array}{llllll}
V &\to &V\,,\qquad  & L_I &\to&L_I - C_{IJK} \gamma^J K^K - \tfrac 12 C_{IJK}\gamma^J \gamma^K V\,,\\[.01\textheight]
K^I &\to &K^I + \gamma^I V\,,\qquad & M&\to &M - \tfrac 12 \gamma^I L_I + \tfrac 1 {12} C_{IJK} \left(V \gamma^I \gamma^J \gamma^K + 3 \gamma^I \gamma^J K^K\right)\,,
\end{array}
\label{gauge}
\ee
where $\gamma^I$ are constants, but leaves the warp factors, rotation vector and field strengths invariant. To reach a patch where there are no Dirac strings at the point $i$, one has to perform a gauge transformation such that the value of $K^I/V$ at this point is zero, or alternatively $K^I$ has no pole. It is not hard to see that to go from a patch where there are no Dirac strings at point $i$ to a patch where there are no Dirac strings at the point $j$, the gauge transformation parameters $\gamma^I$ have to be equal to the flux between the two centers: 
\be
\gamma^I_{ij} = \frac{K^I}{V}\Big|_{r_i}-\frac{K^I}{V}\Big|_{r_j}=  -\Pi^{(I)}_{ij} \,.
\ee

Thus, when changing patch, in order to make sure that one is describing the same supertube, (with the same radius and energy), the {\it effective} supertube charges have to stay the same. Hence the quantized charges have to shift by 
\bea
Q_{1,j} &=& Q_{1,i} + d_3\, \gamma^2_{ij}=   Q_{1,i} - d_3 \Pi^{(2)}_{ij} \,,\nonumber\\
Q_{2,j} &=& Q_{2,i} + d_3\, \gamma^1_{ij}=  Q_{2,i} - d_3 \Pi^{(1)}_{ij} \,,\label{eq:EffChargesShifts}
\eea
where we have denoted by $Q_{1,i}$ and $Q_{2,i}$ the charges of the supertube in the patch where there are no Dirac strings at the point $i$. Note that for supersymmetric minima, the supergravity ``GH charges'' of the backreacted supertubes are equal to the quantized charges \cite{Bena:2008dw}, and the shift of the quantized charges when one changes patch is the same as the shift of the GH charges under the gauge transformation \eqref{gauge}. 

To summarize this discussion, in order to describe the dynamics and vacuum structure of a supertube, one has to work in a fixed patch (in the examples above we have chosen the one where there are no Dirac strings at the point ``1''). However, to understand the physics of a supertube minimum near one of the poles one has to change to a patch where there are no Dirac strings at that pole.

\begin{figure}[ht!]{
\subfigure{
\includegraphics[width=.51\textwidth]{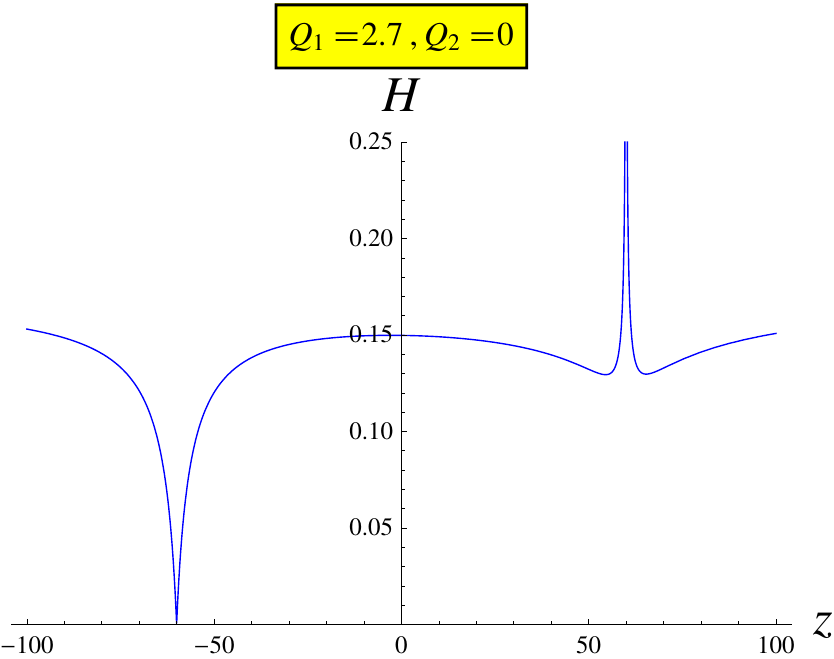}
}
\hspace{0.05\textwidth}
\subfigure{
\includegraphics[width=.39\textwidth]{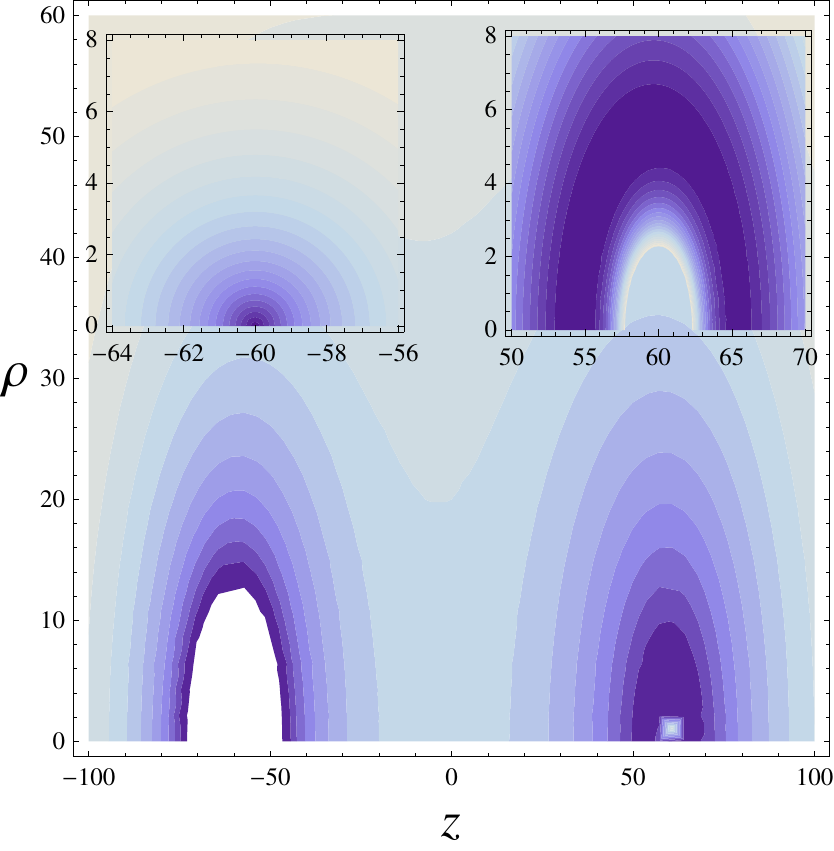}
}
\caption{A metastable minimum and a degenerate supersymmetric at the center on the left. The white region in the contour plot on the right is an artifact of the choice of range for the contour plot; see the insets for a complete picture. The color scales in the insets and main contour plot are different.}\label{fig:metastable_pinch}}
\end{figure}

Armed with the understanding of how to change patches, we are now ready to explain how metastable supertubes decay.
Let us first discuss the example where the metastable supertube decays to a degenerate supersymmetric minimum, illustrated in Figure \eqref{fig:metastable_pinch}.  In the patch with no Dirac strings at the point ``2'', the zero-radius supertube wraps the Dirac string at point ``1''. Hence, this tube is non-contractible (cannot be taken away). In order to reveal that the zero-radius supertube is a bunch of parallel branes, one has to go to the patch with no Dirac strings at point ``1'', and the charges of the branes will shift as in \eqref{eq:EffChargesShifts}. These supersymmetric branes are now parallel to the background, and can be taken away to infinity.

So in the gedanken experiment of lowering the supertube into position at the metastable minimum, having it decay to the supersymmetric minimum and taking the decay products back to infinity, the charge difference between the initial and the final probe branes is
\be
\Delta Q_1 =  -d_3 \Pi^{(2)}_{12}\,, \qquad \Delta Q_2 =  -d_3 \Pi^{(1)}_{12}\,,
\label{chargeshift}
\ee

Furthermore, as we have explained above, when the supertube sweeps out the two-cycle between the points ``$i$'' and ``$j$'' it lowers the $\Pi^{(3)}$ flux of the background by $d_3$ units. Hence, the charges of the background, which come entirely from the magnetic fluxes
\be
Q^{\rm bg}_1 = \Pi^{(2)}_{12} \Pi_{12}^{(3)} \qquad Q^{\rm bg}_2 = \Pi_{12}^{(1)} \Pi_{12}^{(3)}
\ee
are lowered by exactly the amount in \eqref{chargeshift}. The net result of this process is that the negative charges of the metastable tube have annihilated against the positive charges dissolved in flux.

On can repeat the same gedanken experiment with a metastable supertube that decays into a non-degenerate minimum: to take away the decay product one needs again to change patch, and thus shift the charges of the supertube as in \eqref{chargeshift}. The change in charges of the tube is again compensated by a change in the fluxes, which reduces the charges of the background by the same amount.

\begin{figure}[ht!]{ \centering{
\includegraphics[width=0.5\textwidth]{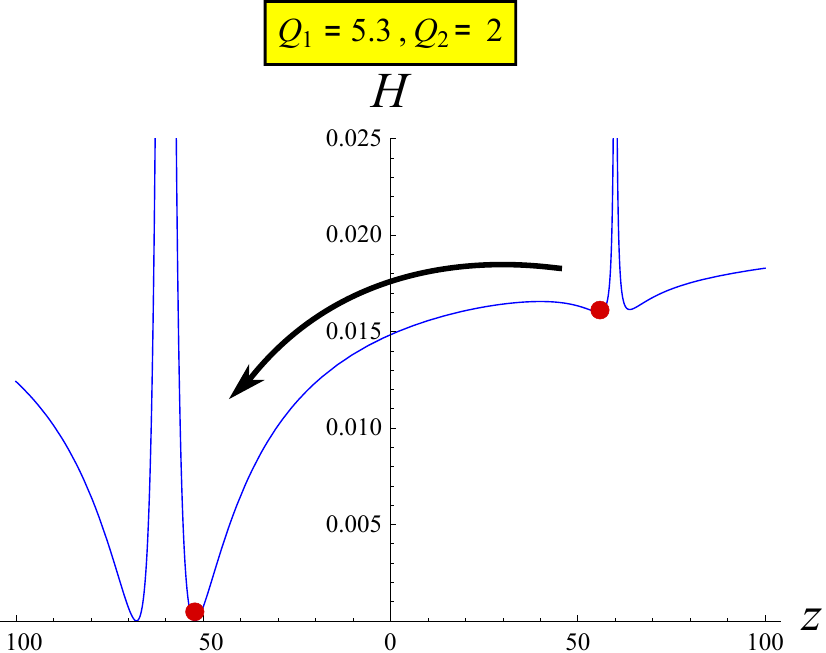}

\vspace{.05\textheight}
\hspace{.03\textwidth}
\includegraphics[width=0.67\textwidth]{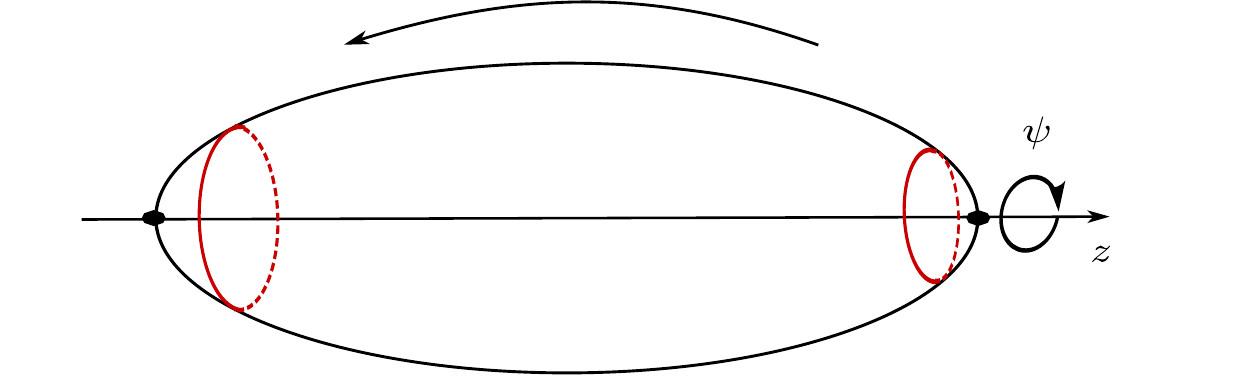}}
\caption{Illustration of the tunneling process. Supertubes are depicted as red circles wrapping the Gibbons-Hawking fiber $\psi$. A metastable supertube close to one center can tunnel to a stable supertube close to the other center, reducing in the process the flux on the two-cycle between these two centers.}
\label{fig:tunnel}}
\end{figure}

\section{Discussion and Future Directions}

We have computed the Hamiltonian of supertubes in three-charge supersymmetric solutions with a Gibbons-Hawking base space, and found that this Hamiltonian can have both supersymmetric and non-supersymmetric minima. In the non-supersymmetric minima one or both of the effective charges of the supertube are oriented opposite to the background charges.

We have then focused on a specific two-center smooth solution, and found that a probe supertube can have also metastable minima, which decay into both supersymmetric and non-supersymmetric stable minima. We have then shown that during the decay, the supertube charges are either partially or totally annihilated against the charges dissolved in flux, much like it happens in other antibrane probe constructions \cite{Kachru:2002gs, Klebanov:2010qs}.

Although we have only illustrated the existence of metastable vacua for a very specific example, we believe it is a generic phenomenon that can occur in all multi-center three-charge backgrounds, and in particular in the smooth solutions with long throats that give microstates of three-charge extremal supersymmetric black holes in five dimensions. The resulting metastable configurations should then represent microstates of non-extremal black holes. It would be interesting to extend our proof-of-concept analysis to such more complicated backgrounds, and to argue that the fuzzballs of extremal black holes survive off-extremality. 

Of course, to fully run this argument, and to understand the properties of non-extremal microstates in the same regime of parameters where the classical black hole exists, one would need to calculate the backreaction of the metastable supertubes, at least to first order. One might argue by analogy with other antibrane backreaction calculations \cite{Bena:2009xk, Bena:2010gs, Bena:2011hz, Giecold:2011gw, Blaback:2011nz} that this may completely wreck the structure of the solution. One counter-argument might be that because our solutions are asymptotically flat, they will be less susceptible to backreaction problems. 

The calculation of the backreacted solution would also reveal whether the non-extremal microstates obtained from metastable supertubes have ergospheres, as one expects \cite{Chowdhury:2008uj} from the JMaRT solution \cite{Jejjala:2005yu}, or have other, more surprising properties.

The other possible issue is that supertube backreaction in a microstate solution with a long throat will drive the throat to become longer, and will cause the fuzzball to collapse behind a horizon. If this happens to the generic microstate, this would imply that the singularity of non-extremal black holes is resolved by configurations that have at least an outer horizon. It is clearly very important to understand the fate of these configurations and whether or not they help to solve the information paradox.

\section*{Acknowledgments}

We thank the participants in the Aspen workshop on ``Singularities and Holography in String Theory'' and the participants in the Benasque workshops on ``String Theory'' and on ``Gravity - New Perspectives from Strings and Higher Dimensions'' for discussions and suggestions, and we are grateful to 
the Aspen Center for Physics and the Centro de Ciencias de Benasque Pedro Pascual for hospitality.  This work was supported in part by the ANR grant 08-JCJC-0001-0,  by the ERC Starting Independent Researcher Grant 240210 - String-QCD-BH as well as by the Aspen Center for Physics NSF Grant 1066293.

\appendix{

\section{Three-charge solutions}
\label{app:three_charge}

\subsection{Three charge solutions in M-theory and type IIA duality frames}

\subsubsection{M-theory solution}\label{sss:M-theory_3Charge}
We begin by reviewing  supersymmetric three-charge solutions in the eleven-dimensional ``M-theory'' duality frame in which the three asymptotic charges of the solution come from M2 branes wrapped on three orthogonal $T^2$'s inside an internal $T^6$ \cite{Bena:2004de}. The ansatz for the metric and three-form is:
\bea
ds_{11}^2 &=& -Z^{-2}(dt + k)^2 + Z \,ds_4^2+Z\sum_{I=1}^3 \frac{ds_I^2}{Z_I}\\
A^{(I)} &=& \sum_{I=1}^3 \left(-Z_I^{-1}{(dt + k)} + B^{(I)}\right) \wedge dT_I\label{eq:11d_solution_app}
\eea
where $Z = (Z_1 Z_2 Z_3)^{1/3}$, $ds_I^2$ and $dT_I$ are, respectively,   a unit metric and unit volume form on the three $T^2$'s and $ds_4^2$ is a four-dimensional hyper-K\"{a}hler metric. The one-form $k$ is supported on this four-dimensional base space and all functions appearing in the solution only depend on the base coordinates. Note that the ansatz for the gauge fields relates the warp functions $Z_I$ appearing in the metric to the electric potentials, the $B^{(I)}$ are magnetic fields on the hyper-K\"{a}hler base.

We are interested in solutions whose four-dimensional base  is with Gibbons-Hawking (GH) or multi-center Taub-NUT
\be
d s^2_4 = V^{-1} (d \psi + A)^2 + V ds_3^2\,,\qquad \star_3 d A = - d V,
\ee
where $V$ is a harmonic function and $ds^3$ is the flat metric on $\mathbb{R}^3$. Solutions with a GH base have a natural interpretation upon KK reduction along the GH fibre $\psi$ (see below), and many supersymmetric bubbling solutions can be constructed \cite{Bena:2005va,Berglund:2005vb,Saxena:2005uk}, for a review see \cite{Bena:2007kg}. See figure \ref{fig:bubblinggeometry} for a cartoon of a bubbling geometry.

The most general supersymmetric 3-charge bubbling solution is determined by 8 harmonic functions $(V,K_I,L^I,M)$ on $\mathbb{R}^3$ which can have an arbitrary number, $n$, of sources:\footnote{We follow the notation of the eleven-dimensional solution of \cite{Bena:2004de}. To make contact to Denef's multi-center solutions in four dimensions \cite{Denef:2000nb,Denef:2002ru,Bates:2003vx}, the dictionary is
\be
V = -\sqrt{2} H^0\,,\qquad K^I = -\sqrt{2} H^I\,, \qquad L_I = \sqrt{2} H_I\,,\qquad M = \frac 1 {\sqrt{2}} H^0\,.\label{eq:H_Denef}
\ee}
\eal{ 
V &= v_0 + \sum_{i = 1}^n \frac{v_{i}}{r_i}\,, \qquad \qquad &M &= m_0 + \sum_{i = 1}^n \frac{m_{0,i}}{r_i}\,,\\
K_I &= k^I_0 + \sum_{i = 1}^n \frac{k^I_{i}}{r_i}\,, &L_I &= \ell_{I,0} + \sum_{i = 1}^n \frac{\ell_{I,i}}{r_i}\,,
}
with the $i$th source sitting at position $\vec r_i$. For the interpretation of $(v,k^I;l_I,m)$ as IIA brane charges, see  table \ref{tab:IIA_Charges}.

The warp factors and magnetic fields of a supersymmetric solution are :
\eqa{
B^{I} &=& V^{-1} K^I (d \psi + A) +  \xi^I \,,\qquad d \xi^I = - \star_3 d K^I \label{eq:Magnetic_field}\\
Z_I &=& L_I + \tfrac 12  C_{IJK}V^{-1} K^J K^K 
}
where $C_{IJK} = |\varepsilon_{IJK}|$. The angular momentum one-form $k$ has the form:\footnote{The rotation vector $\omega$ is a solution to $\star_3 d \omega = \langle H, dH\rangle$ where $\langle\cdot ,\cdot \rangle$ is the antisymmetric symplectic inner product and $H$ the vector of harmonic functions as defined through \eqref{eq:H_Denef}. We do not need the form of $\omega$ in the rest of this paper. }
\eqa{
k &=& \mu (d \psi + A) + \omega,\label{eq:Ang_Mom_k}\\
\mu &=& \tfrac 16 V^{-2} C_{IJK} K^IK^JK^K + \tfrac12 V^{-1} K^I L_I + M.
}

\subsubsection{Two reductions to type-IIA, and their charge interpretatins}

A solution of the form presented above, carries eight types of charges. Six are brane charges in eleven dimensions corresponding to three types of M2 branes, wrapped on three mutually orthogonal $T^2$'s inside $T^6$ (and smeared in the other directions) and three types of M5 branes, wrapped on the dual 4-cycles inside $T^6$ and having one worldvolume direction inside the hyper-K\"{a}hler base. When this base is Gibbons-Hawking, and the M2 and M5 branes respect the GH isometry, one can define
two additional `geometric' charges: (angular) momentum along the GH fiber $\psi$, (controlled by the harmonic function $M$, and Kaluza-Klein monopole charge (controlled by the harmonic function $V$).

When the GH harmonic function $V$ asymptotes to a constant, the GH space becomes multi-center Taub-NUT, which is asymptotically $\mathbb{R}^3 \times S^1$. Upon KK reduction along the $S^1$ the 11-dimensional supergravity solution becomes a four-dimensional CY or torus compactification of type IIA string theory, and the eight charges of the five-dimensional geometry become  asymptotic electric and magnetic charges in four dimensions.

For convenience, in table  \ref{tab:IIA_Charges} we list the interpretation of the eight M-theory charges upon the `standard' reduction to IIA along the GH coordinate $\psi$, and in an alternative reduction to  IIA, over one of the torus directions. The latter, which we call the IIA$'$ frame, is the one we will use in the computation of the supertube Hamiltonian. 

\begin{table}[ht!]
\centering
\begin{tabular}{|ccc||c|c|c|}
\hline
\multicolumn{3}{|c||}{{\bf Charge}}&{\bf M theory}& {\bf IIA: M$/S_{\psi}$} &{\bf IIA$'$: M$/S_1$}\\
\hline
$V$&$\to$&$v_i$&KKm&D6&KKm\\
\hline
$K^I$&$\to$&$ k^1_i$&M5&D4&NS5\\
&&$k^2_i$&M5&D4&NS5\\
&&$k^3_i$&M5&D4&D4\\
\hline
$L_I $&$\to$&$ l_{1,i}$&M2&D2&F1\\
&&$l_{2,i}$&M2&D2&D2\\
&&$l_{3,i}$&M2&D2&D2\\
\hline
$M $&$\to $&$m_i$&$P_\psi$&D0&$P_\psi$\\
\hline
\end{tabular}
\caption{\small Interpretation in the M-theory and two IIA frames we use of the eight charges corresponding to the eight harmonic functions $V,K^I,L_I,M$. $S_\psi$ denotes the GH circle with coordinate $\psi$, $S_1$ one of the directions of torus $T_1$ and $P_\psi$ stands for momentum along the $\psi$ circle (spacetime angular momentum).
 \label{tab:IIA_Charges}}
\end{table}

\subsection{Physical conditions}

There are several consistency conditions GH solutions have to satisfy which translate in part to constraints on the eight constants in the harmonic functions. Fixing the asymptotics of the metric and gauge field further constrains those constants. We also require a physical solution to be free of closed timelike curves (CTC's), by demanding that the metric component $g_{\varphi \varphi}\geq 0$ for any periodically identified direction $\varphi$. If we furthermore impose that the solution has to be smooth, this puts severe restrictions on the allowed set of charges and gives an extra no-CTC condition. We summarize this conditions in the following. (See \cite{Bena:2007kg} for a complete discussion.)

\paragraph{Physical conditions for stable multi-center configurations}

\begin{itemize}
\item Absence of CTC's requires:
\eqn{
Z_1Z_2Z_3 V - \mu^2V^2 \geq0\,,\qquad
V Z_I \geq0\,.\label{eq:CTC_cond}
}
The first condition follows from positivity of  $g_{\psi\psi}$,  the second set is equivalent to having both the polar angle in the three-dimensional base and the $T^6$ directions not to be timelike\footnote{The sufficient no-CTC condition, which insures the existence of a time function is $Z_1Z_2Z_3 V - \mu^2V^2 \geq\ \omega^2$ \cite{Berglund:2005vb}.}.

\item To have a \textit{smooth} geometry, the warp factors and the function $\mu$ appearing in the angular momentum one-form $k$ must be regular as $r_i \to 0$.

This leaves only 4 out of the 8 charges at each center  to be independent. In particular, one finds the relations
\be
l_{I,i} = -\tfrac 12 C_{IJK} \frac{k_i^J k_i^K}{v_i}\,,\qquad m_i = \frac 12 \frac{k_i^1 k_i^2 k_i^3}{q_i^2} \qquad \forall i \text{ (no sum)}\; .\label{eq:regularityharm}
\ee
\item For such a smooth solution, there is a further restriction to ensure absence of CTC's. From the first condition in \eqref{eq:CTC_cond}, namely $Z_1 Z_2 Z_3 V - \mu^2 V^2\geq0$, one notices that $\mu$ has to vanish at each center, since for $r_i\to 0$ the $Z_I$'s tend to finite values while $V^{-1}$ goes to zero:
\be
\mu |_{r_i = 0}= 0\,. \label{eq:smoothbubble}
\ee
By explicitly performing the expansion around each center $\vec r_i$, the latter condition gives $N-1$ so-called \emph{bubble equations} \cite{Bena:2005va,Berglund:2005vb,Bena:2007kg} \footnote{In more general solutions \cite{Denef:2000nb,Denef:2002ru,Bates:2003vx} these equations come from imposing that $\omega$ should have no Dirac-Misner strings at the centers, but in smooth backgrounds this is equivalent to \eqref{eq:smoothbubble}.}
. They relate the magnetic flux (coming from $dB^{I}$) through each bubble to the physical size of each bubble, determined by the inter-center distances $r_{ij}$.\end{itemize}

Depending on the asymptotics, more constraints need to be imposed on the constants $v_0,k^I_0,\ell_{I,0},m_0$. For example, asymptotically $\mathbb{R}^{4,1}$ solutions must have  $v_0 = 0$ and $Z_I \to 1$ at spatial infinity, while asymptotically Taub-NUT solutions have $v_0 \neq 0$.

\section{Hamiltonian for a two-charge tube in a three-charge background}
\label{app:Hamiltonian}

We want to describe two-charge supertubes in three-charge geometries. In the M-theory frame, the two charges of the supertube, $Q_1$ and $Q_2$, correspond to M2 branes wrapped on the two-tori $T_1$ and $T_2$ within the $T^6$. We study whether there are tubular configurations where the two sets of M2 branes blow up into an M5 brane along the GH direction $\psi$.
We denote this tube as \mbox{M2--M2 $\to$ M5}. The method is to write down the Lagrangian (consisting of a Born-Infeld and Wess-Zumino contribution) of an M5 brane with the lower-dimensional charges (corresponding to the two M2 branes) and search for stable configurations. This is done by looking for supersymmetric and non-supersymmetric (meta)stable minima in the Hamiltonian which is obtained from the Lagrangian by a Legendre transform. 

\begin{table}[ht!]
\centering
\begin{tabular}{|c||c|c|c|}
\hline
{\bf Charge}&{\bf M theory}& {\bf IIA: M$/S_{\psi}$} &{\bf IIA$'$: M$/S_6$}\\
\hline
$d_3$&M5&D4&D4\\
$Q_1$&M2&D2&F1\\
$Q_2$&M2&D2&D2\\
$J$&$J_\psi$&D0&$J_\psi$\\
\hline
\end{tabular}
\caption{\small Brane interpretation of the two-charge supertube in a three-charge background. The fourth charge $J$ is related to the others by $|Q_1 Q_2| = |J d_3|$. $J_\psi$ denotes (angular) momentum   along the GH circle $S_\psi$. \label{tab:SupTube_Charges}}
\end{table}

Since the M5 brane worldvolume Lagrangian is rather involved \cite{Pasti:1997gx}, we chose to go to a frame that is more amenable to calculations. In particular, we reduce over one of the torus directions, such that the supertube charges are D2 branes and F1 strings that blow up into a D4 brane, and we will denote it henceforth as a ``D2--F1 $\to$ D4'' supertube. This configuration is analogous (can be seen by two T-dualities) to the original supertube D0--F1 $\to$ D2 \cite{Mateos:2001pi,Emparan:2001ux}: our D2's couple to the magnetic Born-Infeld flux and the F1's to the electric one. For this setup, we know perfectly well how to obtain the Hamiltonian. For a relation of the charges in the M-theory frame to that in the IIA frame, see table \ref{tab:SupTube_Charges}.

\subsection{Reduction of the background along a torus direction}
We work in the IIA$'$ duality frame of table \ref{tab:IIA_Charges}. In the string frame, the NS-NS fields are  \cite{Bena:2008dw}: 
\bea
ds^2_{IIA,st} &=& -(Z_2Z_3)^{-1/2} Z_1^{-1}(dt + k)^2 + (Z_2Z_3 )^{1/2} ds_4^2\\\nonumber\\
&&+  \left(\frac{Z_1 Z_3} {Z_1^2}\right)^{1/2}  d z^2+ \left({Z_2}/{  Z_3}\right)^{1/2}ds_2^2+(Z_3/Z_2)^{1/2} ds_3^2\\
B_2 &=&  A^{(3)} \wedge d x_5\\
e^{4\Phi} &=& { \frac{Z_1 Z_2}{Z_3^2}}
\eea
In the RR sector, the non-trivial fields are $C_3$ and $C_5$. We only list the components we need for computing the supertube Lagrangian:
\bea
C_3 &=&- \left[(Z_2^{-1} -1) dt +(\frac{K^2}V - \frac{\mu}{Z_2})d\psi\right]\wedge dT_2\nonumber\\
C_5&=& -\left[\frac{K^2}{VZ_1} + (\frac{K^1}V - \frac{\mu}{Z_1})\right]dt\wedge d\psi\wedge dz \wedge dT_2\,.
\eea

\subsection{Gibbons-Hawking Hamiltonian}

We consider F1--D2$\to$ D4 tubes with the D4 worldvolume embedding given as
\bea
t=\sigma^0, \qquad \psi=\sigma^1, \qquad z=\sigma^2\,, 
\eea
and $\sigma^3,\sigma^4$ along the torus $T^2$. 

The Lagrangian is
\be
L=-N_{D4}T_{D4} \int \sqrt{g+B_2+\calf_2}\quad + N_{D4}T_{D4} \int (C_5+(B_2+\calf_2)\wedge C_3)\,,
\ee
where ${\calf_2}=2\pi \alpha'F_2$ and $F_2$ is the induced abelian 2-form field strenght on the D4 worldvolume.
The lower-dimensional brane charges are introduced by the worldvolume flux
\be
\calf_2 = \cale d\sigma^0 \wedge d \sigma^2 + \calb d \sigma^3 \wedge d \sigma^4
\ee
The electric field $\cale$ sources IIA string charge along $z$, while the magnetic field $\calb$ induces  D2 brane charge along the torus {$T^2$}.
The supertube Lagrangian (density) becomes
\bea
\call &=& - \frac{d_3} {Z_1 V}\sqrt{ \left[K^3 + V(\calb- \mu(1 -\cale)) \right]^2+ VZ_1Z_2 (1-\cale)(2 - Z_3(1-\cale))}\nonumber\\
&&+ (\frac{1}{Z_1} -1)d_3 \calb+\frac {d_3 K^2}{VZ_1} +d_3 (\frac{\mu}{ Z_1}  + \frac{K^1}{ V}) (1-\cale)\,, \label{eq:Lagrangian}
\eea
with the D4 dipole charge $d_3= N_{D4}T_{D4}$. This Lagrangian was obtained in a different duality frame in \cite{Bena:2008dw} for $d_3 =1$.
The Hamiltonian is obtained by the Legendre transform of $\call$ with respect to the electric field $\cale$,
\be
\calh=\Pi_{\cale} \cale - \call,
\ee
where $\Pi_{\cale}=\frac{\delta \call}{\delta \cale}$ is the momentum conjugate to $\cale$. 

After quite some algebra, one finds the Hamiltonian in terms of the F1 and D2 charges, $Q_1=\Pi_{\cale}$ and $Q_2=d_3 \calb$, and D4 dipole charge $d_3$ 

\eal{
\calh &= \frac{\sqrt{Z_1Z_2Z_3 V^3}}{d_3(Z_1 Z_2Z_3 V - \mu^2 V^2)}\sqrt{\tilde Q_1^2 +d_3^2 \frac{Z_1Z_2Z_3 V - \mu^2 V^2}{Z_2^2 V^2}}\sqrt{\tilde Q_2^2 + d_3^2\frac{Z_1Z_2Z_3 V - \mu^2 V^2}{Z_1^2 V^2}}\\ &
+ \frac{\mu V^2}{d_3(Z_1Z_2Z_3V - \mu^2 V^2)}\tilde Q_1 \tilde Q_2 - \frac 1 {Z_1} \tilde Q_1 - \frac 1 {Z_2} \tilde  Q_2 -  \frac{d_3 \mu}{Z_1Z_2} + Q_1 + Q_2\;, 
}
where we denote $\tilde Q_1 \equiv Q_1+ d_3 (K^2/V - \mu/{Z_2})$ and analogous for $(1\leftrightarrow 2)$ .

\bibliographystyle{toine}
\bibliography{supertube-big}

\end{document}